\documentclass[prb,twocolumn]{revtex4}
\newcommand{\be}{\begin{equation}}
\newcommand{\ee}{\end{equation}}
\newcommand{\ben}{\begin{equation*}}
\newcommand{\een}{\end{equation*}}
\newcommand{\bea}{\begin{eqnarray}}
\newcommand{\eea}{\end{eqnarray}}
\newcommand{\bean}{\begin{eqnarray*}}
\newcommand{\eean}{\end{eqnarray*}}
\newcommand{\nn}{\nonumber}

\usepackage{epsfig}

\begin{document}
\title{Temporal Windowing of Trapped States} \author{L.M.Castellano D.M. Gonzalez}
\affiliation{{\small Instituto de F\'{\i}sica, Universidad de Antioquia, A.A. 1226}\\
Medell\'{\i}n-Colombia}

\begin{abstract}
Trapped state definition for 3-level atoms in $\Lambda$
configuration, is a very restrictive one, and for the case of
unpolarized beams, this definition no longer holds. We introduce a
more general definition by using  a reference frame rotating with
the frequency of the control field, obtaining a temporal windowing
for the trapped population. This amounts to a time quantization of
the coherent population transfer, making possible to study the
phase coherence in trapped light
  PACS number(s): 32.50.+d, 32.80.Qk, 32.80.-t
\end{abstract}\maketitle

\section{Introduction}
\begin{figure}[h]
\centering \epsfig{file=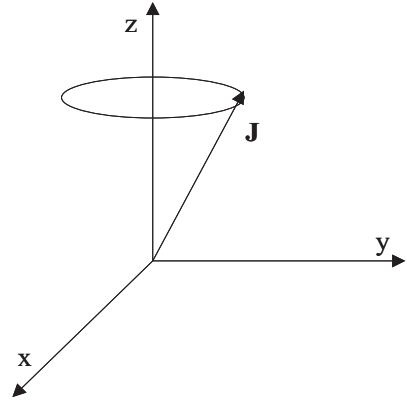, height=150pt, width=150pt}
\caption{\protect \scriptsize The vector polarization
$J\textbf{J}$ rotating with frequency $\Omega$.} \label{jrot}
\end{figure}

3D vector representation had become an important tool in the
handling and understanding of 3-level atoms. News phenomena such
as dark states, electromagnetic induced transparency (EIT)
\cite{Har} , coherent population transfer
(CPT)\cite{sho}\cite{kn}, and the very recent introduction of
polaritons\cite{fle1} to explain the "capture and storing of light
in Cs\cite{lau} and $\sp{87}Rb$\cite{fle2} are easily understood
in the framework of a geometric representation for 3D vector
model. In particular the so named $\Lambda$ configuration with the
two close lying ground states in non allowed Raman transitions had
been largely considered in adiabatic Raman interactions \cite{shk}
and a geometric representation have been proposed. A  usual way of
doing vector models in quantum optics begin with the Maxwell-Bloch
equations for the 2-level atoms. In this case the achieving of a
3D vector representation is immediate with components $ J_x,J_y$
and $J_z$ having a clear physical meaning: the polarization being
$ P = J_x + iJ_y $ and $ J_z $ the population inversion
(FIG.\ref{jrot}). For 3-level atoms similar geometric
representation can be obtained. In this approach, interaction with
two different optical fields are considered (geometrically) as the
sum of two levels atom-field interaction, each field coupling two
different levels independly. However this approach clearly lacks
in rigourously since it is based on the assumption that addition
of two level geometric (the SU(2) Isospin )representation
describes exactly well the three levels atom-field dynamics
(SU(3)Isospin); however it works quite well in 3-level atoms with
$\Lambda$ configuration where the two grounds levels are close
lying.

\section{Basic ideas}
In the following, we briefly review the basic ideas involved in
this approach. The time evolution of vectors like angular momentum
(for example magnetic momentum) is given by:

\begin{equation}  \frac{dJ}{dt} = J \times \Omega.
\label{1}
\end{equation}

This equation describes the precession of vector $J$, around
$\Omega$ axis. A very important fact is that any physical quantity
satisfying EQ.(\ref{1}) it is precesing in space as shown in FIG.
\ref{jrot}. For two level atoms the geometric representation for
the Bloch equations is immediate and comparison with EQ.(\ref{1})
allows us identify the geometric rotation frequencies with
physical quantities: $\Omega_{x}=0 $, $\Omega_{y}=2ga$,
$\Omega_{z} = \triangle$.

\begin{figure}[h]
\centering \epsfig{file=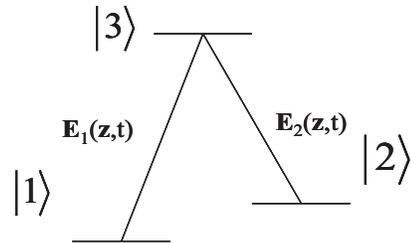 , height=90pt , width=150pt}
\caption{\protect \scriptsize 3level atom in $\Lambda$
configuration with very close lying ground states levels.
$E_1(z,t)$ is the control field and $E_2(z,t)$ is the signal
field} \label{lambda}
\end{figure}

 Following this line of thinking, we write down in what follows,
 the Maxwell-Bloch equations for the the 3-level atoms, in $\Lambda$ configuration (FIG.\ref{lambda} and pursuit
 identical identification as for the 2-level case. The hamiltonian for 3-level atoms
 can be written as
  \begin{eqnarray}
  \mathbf{H} & = &\mathbf{a}_1\mathbf{a}_1^{\dag}\hbar\omega_1+\mathbf{a}_2\mathbf{a}_2^{\dag}\hbar\omega_2\nonumber\\[2mm]
   & + & \mathbf{J}_{22}\hbar\omega_{21}+\mathbf{J}_{33}\hbar\omega_{31}\nonumber\\[2mm]
  & + &g_1(\mathbf{a}_1\mathbf{J}_{21}+\mathbf{a}_1^{\dag}\mathbf{J}_{12})+g_2(\mathbf{a}_2\mathbf{J}_{32}+ \mathbf{a}_2^{\dag}\mathbf{J}_{23})
  \end{eqnarray}

where $g_1=\mu_{13}\sqrt{\frac{\hbar\omega_{1}}{\epsilon_{0}V}}$,
and $g_2= \mu_{23}\sqrt{\frac{\hbar\omega_{23}}{\epsilon_{0}V}}$
represent the atom-field intensity coupling of transitions 1
$\Leftrightarrow$ 3 and 2 $\Leftrightarrow$ 3 respectively, and
$\mathbf{a_1,a_2}$ are the optical fields with frequencies
$w_{1}$, and $w_{2} $ associated with those transitions. As usual
$\mathbf{J_{nm}}$ are the collective atoms raiser ($n>m $) or
lower ($n<m$) operators. The Maxwell-Bloch equations are then
obtained by solving the Heisenberg motion equation
$i\hbar\dot{\mathbf{O}}= [\mathbf{O},\mathbf{H}]$, performing
R.W.A in the $\mathbf{a_1}$ optical field (taking rotating
reference system with $w_1$ frequency) and phenomenological
addition of the decaying terms

\bea \mathbf{\dot{J}_{33}}&=&-\frac{ig_1}{\hbar}(\mathbf{a_1
J_{31}-a_1\sp\dagger J_{13}})-
\frac{ig_2}{\hbar}(\mathbf{a_2 J_{32}-a_2\sp\dagger J_{23}})\nonumber\\
& &-(\Gamma_{13}+\Gamma_{23})\mathbf{J}_{33}\nonumber\\
\mathbf{\dot{J}_{22}}&=&\frac{ig_2}{\hbar}(\mathbf{a_2 J_{32}-a_2\sp\dagger J_{23}})+\Gamma_{23}\mathbf{J}_{33}\nonumber\\
\mathbf{\dot{J}_{11}}&=&\frac{ig_1}{\hbar}(\mathbf{a_1 J_{31}-a_1\sp\dagger J_{13}})+\Gamma_{13}\mathbf{J}_{33}\nonumber\\
\mathbf{\dot{J}_{13}}&=&i\delta_1
\mathbf{J_{13}}-\frac{ig_1}{\hbar}\mathbf{a_1\Delta_{13}}-\frac{ig_2}{\hbar}\mathbf{a_2 J_{12}}-\gamma_{13}\mathbf{J}_{13}\nn\\
\mathbf{\dot{J}_{23}}&=&i(\delta_1+\omega_{21})
\mathbf{J_{23}}-\frac{ig_1}{\hbar}\mathbf{a_1 J_{21}}-\frac{ig_2}{\hbar}\mathbf{a_2\Delta_{23}}-\gamma_{23}\mathbf{J}_{23}\nn\\
\mathbf{\dot{J}_{12}}&=&i\delta_2
\mathbf{J_{12}}+\frac{ig_1}{\hbar}\mathbf{a_1
J_{32}}-\frac{ig_2}{\hbar}\mathbf{a_2\sp\dagger J_{13}}-\gamma_{12}\mathbf{J}_{12}\nn\\
\eea
Where :
$$\mathbf{\Delta_{13}}=\mathbf{J_{11}-{J_{33}}},\ \ \mathbf{\Delta_{23}}=\mathbf{J_{22}-{J_{33}}}$$
$$\delta_{1}=\omega_{1}-\omega_{31},\ \ \delta_{2}=\omega_{1}-\omega_{21}$$
and $\gamma_{ij}$, $\Gamma_{ij}$ are the decaying constants for
polarizations and excitations respectively

In order to get a geometric representation we will assume the
system prepared in a convenient experimental set up as to have a
common $Z$ axis; furthermore we write down for the polarizations
$J_{nm}=J_x^{nm}-iJ_y^{nm}$, if $n>m$ and similarly for the
optical fields $a^{n}= a^{n}_{x}-ia^{n}_{y}$ replacing in equation
(3) and making $g_1=\hbar\ \bar{g}_1$ :

\bea
\dot{\mathbf{J}}_x^{12}&=&\bar{g_2}(\mathbf{a}_2^x\mathbf{J}_y^{13}+\mathbf{a}_2^y\mathbf{J}_x^{13})+\bar{g_1}\mathbf{a}_1^x\mathbf{J}_y^{23}-\delta_2\mathbf{J}_y^{12}-\gamma_{12}\mathbf{J}_x^{12}\nn\\
\dot{\mathbf{J}}_y^{12}&=&\delta_2\mathbf{J}_x^{12}+\bar{g_1}\mathbf{a}_1^x\mathbf{J}_x^{23}-\bar{g_2}(\mathbf{a}_2^x\mathbf{J}_x^{13}-\mathbf{a}_2^y\mathbf{J}_y^{13})-\gamma_{12}\mathbf{J}_y^{12}\nn\\
\dot{\mathbf{J}}_z^{12}&=&2\bar{g_1}\mathbf{a}_1^x\mathbf{J}_y^{13}-2\bar{g_2}(\mathbf{a}_2^x\mathbf{J}_y^{23}+\mathbf{a}_2^y\mathbf{J}_x^{23})\nn\\
& & +(\Gamma_{13}-\Gamma_{23})\mathbf{J}_{33}
\eea

\bea
\dot{\mathbf{J}}_x^{13}&=&\bar{g_2}(\mathbf{a}_2^x\mathbf{J}_y^{12}-\mathbf{a}_2^y\mathbf{J}_x^{12})-\delta_1\mathbf{J}_y^{13}-\gamma_{13}\mathbf{J}_x^{13}\nn\\
\dot{\mathbf{J}}_y^{13}&=&\delta_1\mathbf{J}_x^{13}-\bar{g_1}\mathbf{a}_1^x\Delta_{13}-\bar{g_2}(\mathbf{a}_2^x\mathbf{J}_x^{12}+\mathbf{a}_2^y\mathbf{J}_y^{12})-\gamma_{13}\mathbf{J}_y^{13}\nn\\
\dot{\mathbf{J}}_z^{13}&=&4\bar{g_1}\mathbf{a}_1^x\mathbf{J}_y^{13}+2\bar{g_2}(\mathbf{a}_2^x\mathbf{J}_y^{23}+\mathbf{a}_2^y\mathbf{J}_x^{23})\nn\\
& & +(2\Gamma_{13}+\Gamma_{23})\mathbf{J}_{33}
\eea

\bea
\dot{\mathbf{J}}_x^{23}&=&-(\delta_1+\omega_{21})\mathbf{J}_y^{23}-\bar{g_1}\mathbf{a}_1^x\mathbf{J}_y^{12}-\bar{g_2}\mathbf{a}_2^y\mathbf{\Delta}_{23}-\gamma_{23}\mathbf{J}_x^{23}\nn\\
\dot{\mathbf{J}}_y^{23}&=&(\delta_1+\omega_{21})\mathbf{J}_x^{23}-\bar{g_1}\mathbf{a}_1^x\mathbf{J}_x^{12}-\bar{g_2}\mathbf{a}_2^x\mathbf{\Delta}_{23}-\gamma_{23}\mathbf{J}_y^{23}\nn\\
\dot{\mathbf{J}}_z^{23}&=&2\bar{g_1}\mathbf{a}_1^x\mathbf{J}_y^{13}+4\bar{g_2}(\mathbf{a}_2^x\mathbf{J}_y^{23}+\mathbf{a}_2^y\mathbf{J}_x^{23})\nn\\
& & +(\Gamma_{13}+2\Gamma_{23})\mathbf{J}_{33}
\eea

\subsection*{The construction of vector $\mathbf{J}$}

The construction of a geometric vector representation is done by
considering $\omega_{21}<<\omega_{31}, \omega_{23}$. This mean
that in the absence of any optical field coupling the transition
$1 \Leftrightarrow 2$, the contribution of this transition in the
absorbtion of photons from the optical fields, which is due to
imaginary part $\mathbf{J}_{y}^{12}$ is neglected and the same can
be argued for the diffractive part $\mathbf{J}_{x}^{12}$. We could
say then that in the overall dynamic of the vector polarization
$\mathbf{J}$, the influence of $\mathbf{J}_{12}$ is negligible. we
do not need to consider  the role of spontaneous emission since
for any dark state the relaxation rate  $\Gamma$ is very
small\cite{hus}(below 3kHz for sodium). With this approach EQ. (5)
and (6) are now:

\bea
\dot{\mathbf{J}}_x^{13}&=&-\delta_1\mathbf{J}_y^{13}\nn\\
\dot{\mathbf{J}}_y^{13}&=&\delta_1\mathbf{J}_x^{13}-\bar{g_1}\mathbf{a}_1^x\Delta_{13}\nn\\
\dot{\mathbf{J}}_z^{13}&=&4\bar{g_1}\mathbf{a}_1^x\mathbf{J}_y^{13}+
2\bar{g_2}(\mathbf{a}_2^x\mathbf{J}_y^{23}+\mathbf{a}_2^y\mathbf{J}_x^{23})
\eea

\bea
\dot{\mathbf{J}}_x^{23}&=&-(\delta_1+\omega_{21})\mathbf{J}_y^{23}-\bar{g_2}\mathbf{a}_2^y\mathbf{\Delta}_{23}\nn\\
\dot{\mathbf{J}}_y^{23}&=&(\delta_1+\omega_{21})\mathbf{J}_x^{23}-\bar{g_2}\mathbf{a}_2^x\mathbf{\Delta}_{23}\nn\\
\dot{\mathbf{J}}_z^{23}&=&2\bar{g_1}\mathbf{a}_1^x\mathbf{J}_y^{13}+4\bar{g}_2(\mathbf{a}_2^x\mathbf{J}_y^{23}+\mathbf{a}_2^y\mathbf{J}_x^{23})
\eea Identification of Rabbi frequencies are immediate for each
transition: $\Omega_{z}^{13}=\delta_{1}$,
$\Omega_{x}^{13}=4\bar{g}_1\mathbf{a}_{1}^{x}$ and
$\Omega_{y}^{13}=0$, for the $1 \Leftrightarrow 3$ and
$\Omega_{z}^{23}=(\delta_{1}+\omega_{21})$,
$\Omega_{x}^{23}=4\bar{g_2}\mathbf{a}_{2}^{x}$,
$\Omega_{y}^{23}=4\bar{g_2}\mathbf{a}_{2}^{y}$ for the $2
\Leftrightarrow 3$ transition. The resulting (7) and (8)
equations, are easily interpreted geometrically recognizing: \bea
\Omega_{x}^{23}&=&4\bar{g_2}\mathbf{a}_2\cos\Delta t \nn\\
\Omega_{y}^{23}&=&4\bar{g_2}\mathbf{a}_2\sin\Delta t\eea where we
have defined the field detunings $\Delta = \omega_{1}-\omega_{2}$.

\begin{figure}[h] \centering
\epsfig{file=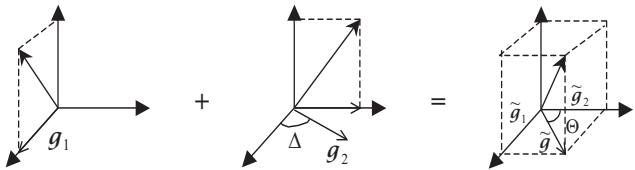 , height=70pt , width=240pt}
\caption{\protect \scriptsize The FIG. at the right hand side
represents the polarization of 3-level atom considered as addition
of 2level atoms polarization when interaction $1\Leftrightarrow2$
is neglected} \label{suma}
\end{figure}

Since R.W.A have been made choosing a reference frame system which
is rotating with the $\omega_{1}$ frequency,the field $a_{2}$ it
appears as rotating with $\Delta$ frequency, we also have choose,
as usual, the imaginary part of $a_{1}$ in the Rotating system
equal to cero. The FIG. \ref{suma} represent the geometric
realization of the whole dynamic for the 3-level atoms in the
framework of non allowed raman transitions (or very close-lying
levels)

\section{Redefining Trapped States }

The customary trapped state definition is given as \bea
\mid->&=&\cos\theta\mid1>-\sin\theta\mid2>\nn\\
\mid+>&=&\sin\theta\mid1>+\cos\theta\mid2> \eea with $\sin\theta =
\frac{g_1}{g}$ and $\cos\theta = \frac{g_2}{g}$, where $g =
\sqrt{g_1^{2}+g_{2}^{2}}$.

Definition (10) corresponds to a similarity transformations around
Z axis. Clearly, the geometric interpretation can be taken as a
summation of the two isospin realization for the case of two
levels atoms, with the two fields falling orthogonal to the atom
and no coupling between levels $1\Leftrightarrow 2$. We find this
definition a very restrictive one, despite to the possibility of
its experimental realization. We redefine

\bea \tilde{g}_1 &=&g_1+g_2\cos\Delta t\nn\\
     \tilde{g}_2 &=&g_2\cos\Delta t\nn\\
     \tilde{g}&=&\sqrt{\tilde{g}_1^{2}+\tilde{g}_2^{2}}
     \eea
     and hence
     \bea
     \sin\Theta(t)&=&\frac{\tilde{g}_1}{\tilde{g}}\nn\\
     \\cos\Theta(t)&=&\frac{\tilde{g_2}}{\tilde{g}}
     \eea

With this redefinition \be <3\mid H
\mid->=\frac{\hbar}{\tilde{g}}(g_1\tilde{g}_2-g_2\tilde{g}_1) \ee

Since we are looking for trapped states then we required $ <3\mid
H^{'}\mid->=0$, and therefore \bea
g_1\tilde{g}_{2}&=&g_{2}\tilde{g}_{1}\nn\\
g_{2}\sin\Delta t&=&g_{1}(1-\cos\Delta t)\nn\\
(\frac{g_{2}}{g_{1}})^{2}&=&\frac{1-\cos\Delta t}{1+\cos\Delta t
}\eea

Equation (15) is a key result in this work. From here we are able
to get some interesting results depending on the relative coupling
intensity. In general the optical field $\mathbf{a}_1$ is a
control field usually more intense than $\mathbf{a}_2$ which is
considered a signal field. We investigate two particular
situations:

1)case $\frac{g_2}{g_1}\simeq 0$. For this situation we obtain
\be\Delta t = 2n\pi \ee, where $n = 0,1,2,...$ this implies a
quantization of the time for which the excited state is a dark
state or trapped state according to \be t =
\frac{2n\pi}{\Delta}\ee.

For a giving optical detunings the temporal windows for achieving
trapped states are $\frac{1}{\triangle\nu}$,
$\frac{2}{\triangle\nu}$, $\frac{3}{\triangle\nu}$.. etc, where
$\triangle\nu = \nu_{1}-\nu_{2}$

2)case $g_{2}\simeq g_{1}$, then \be \Delta t =
(n+\frac{1}{2}\pi)\ee where $n = 1,2,3,...$ and the length of the
quantization times are now
$\frac{3}{4\triangle\nu}$,$\frac{5}{4\triangle\nu}$,$\frac{7}{4\triangle\nu}$...etc

We can see that in both cases this length it depends critically on
the optical detunings and for the case of resonance becomes
infinite and continue. This mechanism allows the storages and
release of the atomic population in each temporal windows, making
possible the study of spin wave interference by using the properly
intensity of control and signal fields,  shifted in time according
to the former result. We point out that study of polaritons
properties can be made using this mechanism.

\section{Conclusions}
We have shown that for the case of unpolarized optical fields
coupling transitions in 3-level atoms with $\Lambda$
configuration, and close-lying ground states levels,the capture of
atomic population in a trapped state it is time windowed. For a
giving detunings $\triangle\nu$ this window is discrete phase
shifted and critically  depending on the relation of intensity
coupling of control and signal field. It comes out to our
attention the recent report\cite{exp} of experimental evidence of
phase coherence. In this report a technic of pulsed magnetic field
is used to vary the phase of atomic spin excitation which are
converted in light and then, throughout interference, detects
phase difference. In this case the pulsed magnetic field induce
the " temporal windowing" for the population trapping.

\section{Acknowledgments}

We are thankful to professor Jorge Mahecha for helpful discussions
and comments. D.M. Gonzalez wants to thanks MAZDA  Foundation in
Colombia for its economical sponsor. This work had been realized
upon grant INF71C of the Centro de Investigaciones Exactas y
Naturales of Universidad de Antioquia.

\end{document}